\documentclass[aps,twocolumn,superscriptaddress]{revtex4}
\usepackage{epsfig}
\usepackage{times}
\begin{document}

\title{Chimera states in uncoupled neurons induced by a multilayer structure}

\author{Soumen Majhi}
\affiliation{Physics and Applied Mathematics Unit, Indian Statistical Institute, Kolkata-700108, India}

\author{Matja{\v z} Perc}
\email{matjaz.perc@uni-mb.si}
\affiliation{Faculty of Natural Sciences and Mathematics, University of Maribor, Koro{\v s}ka cesta 160, SI-2000 Maribor, Slovenia}
\affiliation{CAMTP -- Center for Applied Mathematics and Theoretical Physics, University of Maribor, Krekova 2, SI-2000 Maribor, Slovenia}

\author{Dibakar Ghosh}
\affiliation{Physics and Applied Mathematics Unit, Indian Statistical Institute, Kolkata-700108, India}

\begin{abstract}
Spatial coexistence of coherent and incoherent dynamics in network of coupled oscillators is called a chimera state. We study such chimera states in a network of neurons without any direct interactions but connected through another medium of neurons, forming a multilayer structure. The upper layer is thus made up of uncoupled neurons and the lower layer plays the role of a medium through which the neurons in the upper layer share information among each other. Hindmarsh-Rose neurons with square wave bursting dynamics are considered as nodes in both layers. In addition, we also discuss the existence of chimera states in presence of inter layer heterogeneity. The neurons in the bottom layer are globally connected through electrical synapses, while across the two layers chemical synapses are formed. According to our research, the competing effects of these two types of synapses can lead to chimera states in the upper layer of uncoupled neurons. Remarkably, we find a density-dependent threshold for the emergence of chimera states in uncoupled neurons, similar to the quorum sensing transition to a synchronized state. Finally, we examine the impact of both homogeneous and heterogeneous inter-layer information transmission delays on the observed chimera states over a wide parameter space.
\end{abstract}

\maketitle

The interaction among coupled oscillators in a system often results in fascinating spatiotemporal patterns and one of the most surprising among them is the \emph{chimera state} which consists of coexisting domains of spatially coherent (synchronized) and incoherent (desynchronized) oscillators. After observing firstly in a nonlocally coupled system of identical phase oscillators \cite{chp1}, chimera states have been extensively studied during the past decade. They have been observed in a wide range of systems, for example in phase oscillators \cite{chp1,chp2,chp3,chp4,chp5,chp6}, neuronal models \cite{chne1,chne2,chne3,chne4,chne5,chne6,chne7}, chaotic systems \cite{chco1,chco2}, Hopf normal forms \cite{chhn1,chhn2,chhn3,chhn4} etc. Even the coupling topology is not restricted to the nonlocal one, rather chimera states have been noticed in global \cite{global,global_a,global_b} as well as in local interactions \cite{chne5,chne6,chlo1} and even for one-way local coupling \cite{tanmoy}.  Chimera-like states have also been well investigated on complex networks \cite{chcom}, time-varying network \cite{chtv} and modular type neural networks \cite{chne6}. Apart from numerical and theoretical studies, experimental evidence of chimera states is reported in optical coupled-map lattices \cite{chexp1}, coupled chemical oscillators \cite{chexp2}, metronomes \cite{chexp3} and squid meta-materials \cite{chexp4} etc.

The existence of chimera state is strongly connected to neuronal systems, e.g. various types of brain diseases \cite{brain_disease1,brain_disease2} such as Parkinson's disease, epileptic seizures, Alzheimer's disease, schizophrenia and brain tumors. Chimera states are also related with the real world phenomena of unihemispheric slow-wave sleep \cite{sleep} of some aquatic animals (e.g. dolphins) and migrated birds. During slow-wave sleep in these species half part of the brain is in sleep and the other half remains awake. This strongly indicates that the neurons of the sleepy part are synchronized (coherent) and desynchronized (incoherent) in the awake part of the cerebral hemisphere, which resembles the chimera state. Recently, chimera states are observed in neuronal oscillators if the neurons are locally, globally or nonlocally coupled via chemical synapses \cite{chne5}. But what if the neurons are not connected to each other?  Diverse nervous
activities are found not only among coupled
neuron groups in the same brain region, but also among
uncoupled neuron groups in the same brain region or among
different cortical areas. It can even cross over two
semi-spheres of the brain. Thus in the nervous system, activities are present not only among the coupled neurons, but also among the uncoupled neurons.
Studies on neuron synchronization are mainly focused
on two cases: the coupled neurons and the uncoupled
neurons. Previously, synchronization is observed in uncoupled neurons subject to a common noisy field \cite{noise} and also under neuron's membrane potential stimulation \cite{uncsync} . Experimentally it was observed that synchronization in different neurons may appear in the same region of the brain and even in different regions in the brain \cite{brain}.  On the other hand, most of the earlier works on chimera states assume that the oscillators are on a single layer whatever be the network topology is. Again, there are many physical systems that do not interact directly but exchange the information through a common medium, for instance, in the Huygen's experiment the two pendulum clocks were interacting through the common wooden beam from which they were hanging. Also this type of indirect interaction is particularly important in biological systems, e.g., populations of cells in which oscillatory reactions are taking place \cite{envi} through the interaction via chemicals that diffuse in the surrounding medium.

In this paper, we investigate an architecture where the neuronal oscillators in the upper layer (the layer of our interest) have no connection among them, while they interact with each other via another layer of similar oscillators, thus forming a multilayer structure. To the best of our knowledge, this is the first observation of chimera states in uncoupled neurons in the form of multilayer network. Recent research and reviews attest to the fact that multilayer networks are the next frontier in network science \cite{buldyrev_n10, gao_jx_np12, havlin_pst12, helbing_n13, gomez_prl13, radicchi_np13, de2013mathematical, nicosia_prl13, kivela_jcn14, majdandzic2014spontaneous, podobnik2014network, boccaletti_pr14, podobnik2015cost, wang2015evolutionary, dibaRussia}. In particular, not only are the interactions between the constituents of a complex system limited and thus best described by networks, it is also a fact that the processes happening in one network may vitally affect another network, and moreover, that a node in one network is likely part of another network. From the world economy and transportation systems to social media and biological systems, it is clear that such interdependencies exist, thus making networks of networks or multilayer networks a much better and realistic description of such systems.

This type of multilayer structure is also quite evident in neuronal networks. So it would be of obvious importance to study a network having an architecture of the this type taking neuronal models as the nodes of both the layers. We take the neuron dynamics in terms of Hindmarsh-Rose neuronal model which exhibits various types of bursting dynamics such as spiking, plateau bursting, square-wave bursting (periodic and chaotic) and mixed mode bursting etc. depending on the system parameters. Two types of synapses i.e. electrical and chemical synapses exist through which neurons communicate with each others. Moreover, neurons may not be connected with each other with the same type of synapses everywhere. In fact, a recent work on chimera-like states has been done on neural network inspired by the neuronal connection of the C. elegans soil worm, organized into six interconnected communities, assuming that the neurons are connected with electrical synapses within the communities and with chemical synapses across them \cite{chne6}. In this study, we consider that the neurons are connected with electrical synapses within the bottom layer (the medium) and with chemical synapses across the layers.  Again in neuronal networks, delays arise due to finite propagation times along the axons or to reaction times at chemical synapses. Also depending on the physiological properties of axons and synapses, these delays in signal transmission between different cells of the network may differ. Thus time delay in inter-layer information transmission process is indisputable and it can be heterogeneous too. Hence our aim in this article is to examine how these two types of synaptic connection affect the dynamics of upper layer and how chimera pattern emerges due to that competing effects of two synapses and finally to study the influence of inter-layer synaptic delay (both homogeneous and heterogeneous) on the upper layer dynamics.

\begin{figure}
\centerline{\epsfig{file=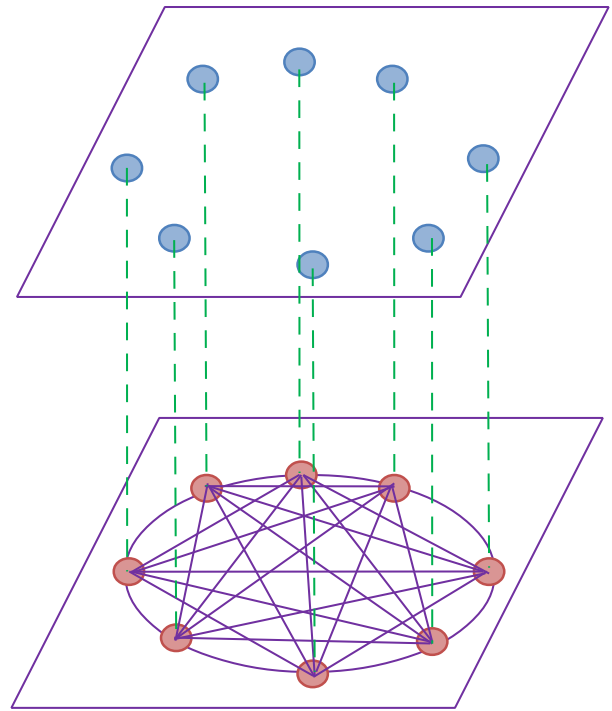,width=8cm}}
\caption{Schematic diagram of a multilayer network where the neurons in the upper layer (blue circle) are uncoupled, while the neurons in the lower layer (red circle) are globally (all-to-all) coupled through electrical synapses (solid lines). Connections  exist between the corresponding neurons in the lower and upper layer through chemical synapses (dashed lines). Each neuron in the upper layer is connected to its immediate bottom neuron in lower layer.}
\label{schematic}
\end{figure}

We consider $N$ identical isolated neurons which are connected through a common medium. We assume that isolated neurons are situated on same layer (upper layer) and medium as globally connected neurons (lower layer). Each isolated neuron interact directly with one neuron (its replica) in the medium. As we are considering global coupling in the multi-layering layer (common medium) and the uncoupled neurons are only interacting with its replica in the common medium, so the spatial order of the neurons in the upper layer is  same as the spatial order of the neurons in the common medium. The schematic diagram of the network is shown in Fig.~\ref{schematic}.  Therefore, in terms of  recently developed multi-layer networks, our proposed network is a multilayer network with two layers.  We consider each neuron as Hindmarsh-Rose neuron model. The neurons in the medium are connected through electrical synapses as they allow direct and passive flow of electron via gap junctions. These gap junctions permit for mutual instantaneous transmission of electron between the neurons which are spatially very close to each other. The main goal for considering electrical synapses in the lower layer (medium) is to synchronize (or coherent motion) electrical activity among the neurons. We assume electrical synapses among the globally coupled neurons in the common medium which is homogeneous distribution of the medium. This assumption is more realistic in biological and chemical systems, as homogeneous medium is observed either in biological systems by the fast diffusion of the small molecules or in chemical systems by stirring the solution. On the other hand, the isolated neurons are connected with the common medium through chemical synapses. The chemical synapses typically function in a longer range compared to electrical synapses, it would thus be more likely to connect across the layers. Thus the simultaneous effect of electrical and chemical synaptic coupling is best represented by multilayer structure where the neurons in the medium are connected through electrical synapses while across the layers through chemical synapses.

\section*{Results}

\begin{figure*}
\centerline{\epsfig{file=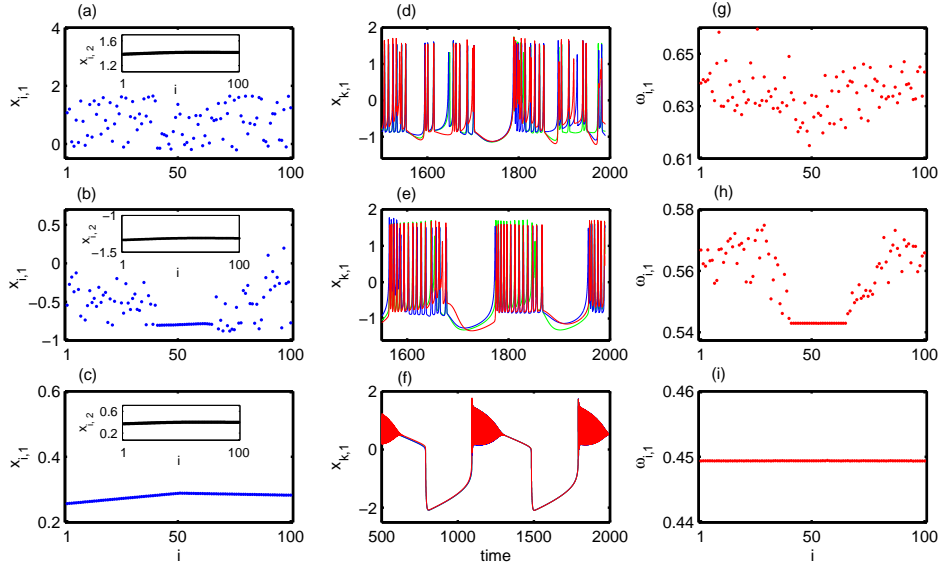,width=15cm}}
\caption{Left panels show snapshots of membrane potentials $x_{i, 1} (i=1,2,...,100)$ for (a) disordered state, $K_{ch} = 1.0$, (b) chimera state, $K_{ch} = 1.130$, and (c) coherent state, $K_{ch} = 1.30$. Middle panels show the behaviors of the neurons $x_{k, 1} (k=20, 50, 90)$ in the upper layer for (d) disordered, (e) chimera, and (f) coherent states. Right panels show corresponding mean angular frequencies $\omega_{i, 1} (i=1,2,...,100)$ for (g) disordered, (h) chimera and (i) coherent states. The inset figures in (a), (b) and (c) are the corresponding snapshots of the lower layer neurons' membrane potentials $x_{i, 2} (i=1,2,...,100)$ (black color), that signify coherent state of the neurons.}
\label{snapshot_nodelay}
\end{figure*}

In this work, we study the emergence of symmetry breaking pattern in the upper layer as a result of the co-action of the two types of synapses.  Here we are considering multilayer network and the number of neurons in both the layers are same. We investigate two different cases based on inter-layer coupling delays. In first case, we consider the instantaneous inter-layer chemical synaptic coupling between the layers and later the effect of delays (homogeneous and heterogeneous) present in the inter-layer chemical synaptic coupling.

\subsection*{Instantaneous inter-layer chemical synaptic interaction}

In this section, we mainly investigate the dynamics of the isolated neurons in absence of inter-layer chemical synaptic coupling delay. For $K_{ch}=0$ (in Eq.(4) of Method section), the bottom layer gets synchronized quite rapidly due to the global (all-to-all type) interaction between the neurons with electrical synapses (particularly, $K_{el}=1.0$) and the neurons in the upper-layer behave according to their individual rhythms (i.e., the rhythm of square wave bursting). So, our objective is to explore the dynamics of the uncoupled neurons in the upper layer while activating the inter-layer chemical synaptic coupling strength  $K_{ch}$ and keeping the bottom layer neurons synchronized. In this case the uncoupled neurons are connected with the common medium.   Now switching on $K_{ch}$, we initially see the incoherent (desynchronized) behaviors of the upper layer neurons and they remain incoherent for $0\le K_{ch}< 1.075$ , but as we increase $K_{ch}$, the upper layer network spontaneously splits into two coexisting domains, one of which is coherent and the other one is incoherent which portrays the feature of \emph{chimera states}. If we increase $K_{ch}$ further, we observe that all the neurons get synchronized for $K_{ch}>1.230$. Figures~\ref{snapshot_nodelay}(a,b,c) show the snapshots of membrane potentials of all the uncoupled neurons in the upper layer exhibiting incoherent, chimera and coherent states at $K_{ch}=1.0, 1.13$ and $1.30$ respectively. At these points, the coherent behavior of all the neurons in the common medium are illustrated in insets of Figs.~\ref{snapshot_nodelay}(a,b,c). Middle panel of Fig.~\ref{snapshot_nodelay} shows the behaviors of particular three neurons $x_{20, 1}, x_{50, 1}$ and $x_{90, 1}$ shown by green, blue and red colors respectively. At incoherent states, all the upper layer neurons follow chaotic square-wave bursting dynamics (Fig.~\ref{snapshot_nodelay}(d)). At higher value of inter-layer chemical synaptic coupling $K_{ch}=1.13$, a typical pattern of chimera state with one group of coherent neurons and another group of incoherent neurons coexist. In this case, a neuron in the coherent group and a neuron in the incoherent group have the same time series form i.e. chaotic square-wave bursting in nature shown in Fig.~\ref{snapshot_nodelay}(e). At higher value of $k_{ch}=1.3$, all the uncoupled neurons in the upper layer are found to be in coherent state and the neurons are in plateau bursting states (Fig.~\ref{snapshot_nodelay}(f)).  Additionally, we calculate mean angular frequency \cite{maf} of the $i$-th neuron as,
$$\omega_{i, 1}=\left \langle \dot{\phi}_{i, 1} \right \rangle_t=\frac{x_{i, 1} \dot{y}_{i, 1}-\dot{x}_{i, 1} y_{i, 1}}{x_{i, 1}^2+y_{i, 1}^2},$$
where $\phi_{i, 1}=\mbox{arctan}(y_{i, 1}/x_{i, 1})$ is the geometric phase for the fast variables $x_{i, 1}$ and $y_{i, 1}$ of the $i$-th neuron, which is a good approximation as long as $c$ is small ($<<1$) and $\left \langle...\right \rangle_t$ denotes long term time average. The mean angular frequencies corresponding to incoherent, chimera and coherent states are shown in Figs.~\ref{snapshot_nodelay}(g, h, i) respectively. To calculate mean angular frequencies $\omega_{i, 1}$, the time interval is taken over $5 \times 10^5$  time units after an initial transient of $3 \times 10^5$ units, throughout the paper. The mean angular frequency corresponding to the neurons in incoherent group are randomly scattered whereas for coherent group of neurons they are same.  These  mean angular frequency profiles clearly distinguished coherent and incoherent groups in the chimera state.

\begin{figure}
\centerline{\epsfig{file=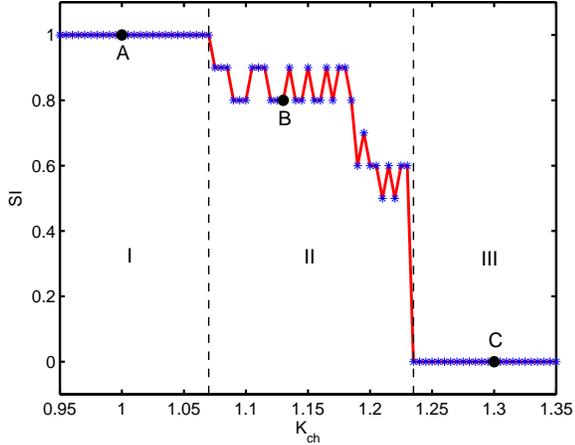,width=8.5cm}}
\caption{Variation of strength of incoherence (SI) is plotted by changing inter-layer chemical synaptic coupling strength $K_{ch}$. Regions I, II and III are respectively for incoherent, chimera and coherent states. Here $N = 100$, $M = 20$, and $ \delta = 0.05$. The values pointed as A (at $K_{ch}=1.0$), B (at $K_{ch}=1.130$) and C (at $K_{ch}=1.30$) respectively correspond to the exemplary snapshots shown in Fig.~\ref{snapshot_nodelay}(a), (b) and (c) respectively.}
\label{SIvary}
\end{figure}

Figure~\ref{SIvary} depicts the variation of strength of incoherence (SI) (refer to the Method section) with respect to inter-layer chemical synaptic coupling strength $K_{ch}$. As can be seen, in the region I=$\{K_{ch}: 1.0\le K_{ch}<1.075\}$, the value of SI remains unity characterizing the incoherent (disordered) neurons but as we increase $K_{ch}$ beyond $K_{ch}=1.075$, we observe chimera state characterized by the values $0<$ SI $<1$ in the region II=$\{K_{ch}: 1.075\le K_{ch}\le 1.230\}$. Although $0<$ SI $<1$ may represent other dynamical states like cluster state \cite{SImeasure}, splay state etc., rigorous verification of the snapshots and time series of the neurons in the parameter range II (also for all the parameter region plots in the following sections) have confirmed the existence of chimera patterns only. With further increase, the value of $K_{ch}$ leads to the coherent state as the values of SI becomes zero in the region III=$\{K_{ch}: K_{ch}> 1.230\}$.

\subsection*{Quorum sensing mechanism for chimera states}

\begin{figure*}
\centerline{\epsfig{file=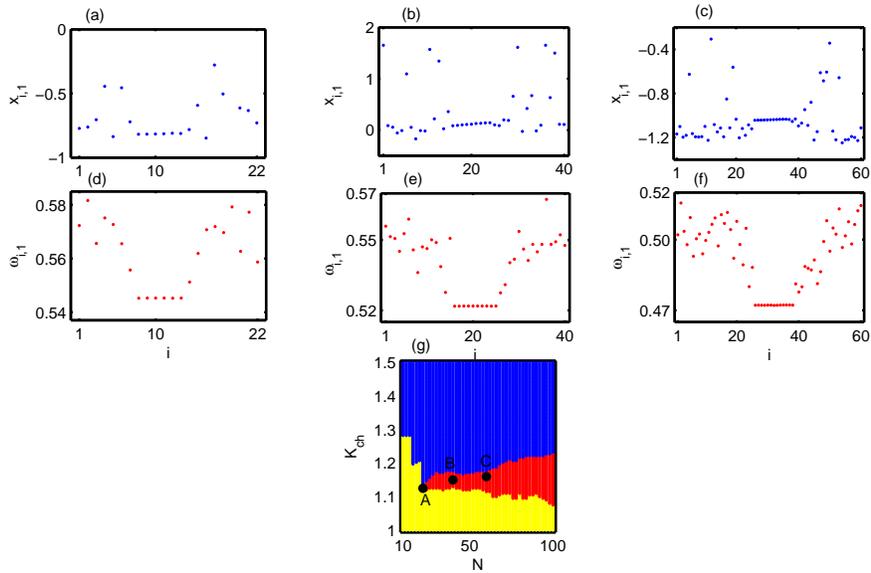,width=14.8cm}}
\caption{Snapshots illustrating emergence of chimera states for (a) $N=22, K_{ch}=1.125$, (b) $N=40, K_{ch}=1.15$ and (c) $N=60, K_{ch}=1.16$. Lower panels (d, e, f) show corresponding mean angular frequencies for $N=22, 40$ and $60$ respectively. (g) Phase space diagram in inter-layer chemical synaptic coupling strength $K_{ch}$ and the number of uncoupled neurons $N$. The region of disordered, chimera and coherent states are represented by yellow, red and blue colors respectively. Strength of incoherence is used to distinguish different states. The points A, B and C correspond to the values used in (a,d), (b,e) and (c,f) respectively.}
\label{quorum}
\end{figure*}

Next, we find the density-dependent threshold for the emergence of chimera states in the upper-layer. This density-dependent threshold is a similar entity like that in quorum-sensing  transition to synchronization \cite{quorum0,quorum}. This mechanism plays a key role in bacterial infection, biofilm formation and bioluminescence \cite{bacteria}. In the context of neuronal network, a similar quorum sensing mechanism involve local field potential \cite{bio1,bio2} which may exist through a different level in cortical hierarchy and play an important role in the synchronization of group of neurons.  This mechanism also exists in the synchronization of chemical oscillators \cite{chem1} and cold atoms \cite{cold}.  In fact, many natural synchronization phenomena where the individual oscillators are not directly coupled, but coupled rather through a common medium experience  different synchronization regimes as a function of the number of uncoupled nodes or their density. In our work, by increasing the number of uncoupled neurons in the upper layer (as well as the number of neurons in the lower layer) which are interacting through lower layer (medium), an emergence of chimera states is observed in the upper layer. For small number of uncoupled neurons, say $N=22,$ we observe chimera states for very small range of $K_{ch} (1.12\le K_{ch} \le 1.13)$ and at higher value of it gives coherent state. Chimera states are not identified for $N<22$ and in this case all the neurons are either in disordered or coherent state depending on the value of $K_{ch}.$ The snapshot of the membrane potential $x_{i, 1}$ for $N=22$ and $K_{ch}=1.125$ is shown in Fig.~\ref{quorum}(a). If we increase the number of uncoupled neurons in the upper layer, the range of $K_{ch}$ increases for the existence of chimera states. For $N=40,$ chimera emerges for the inter-layer chemical synaptic coupling strength $K_{ch}$ in $(1.13\le K_{ch} \le 1.175)$ and snapshot of $x_{i,1}$ at $K_{ch}=1.15$ are shown in Fig.~\ref{quorum}(b). Figure~\ref{quorum}(c) shows the chimera states for $N=60$ and $K_{ch}=1.16$. Figures~\ref{quorum}(d, e, f) show the mean angular frequencies $\omega_{i, 1}$ corresponding to the chimera states in Figs.~\ref{quorum}(a, b, c) respectively. To calculate mean angular frequencies $\omega_{i, 1}$, the time interval is taken over $5 \times 10^5$ time units after an initial transient of $3 \times 10^5$. It is observed that the chimera states persist for long time range in the case of small number of uncoupled neurons as well. To explore the complete dynamics of the uncoupled neurons, we plot the phase diagram in the $N-K_{ch}$ parameter
 space (Fig.~\ref{quorum}(g)) for the range of $N\in [10, 100]$ and $K_{ch} \in[1.0, 1.5]$.  From this figure, it is seen that the region of chimera states in the inter-layer synaptic coupling strength $K_{ch}$ increases with  increasing density $N$ of the neurons.

\subsection*{Homogeneous inter-layer chemical synaptic coupling delay}

\begin{figure*}
\centerline{\epsfig{file=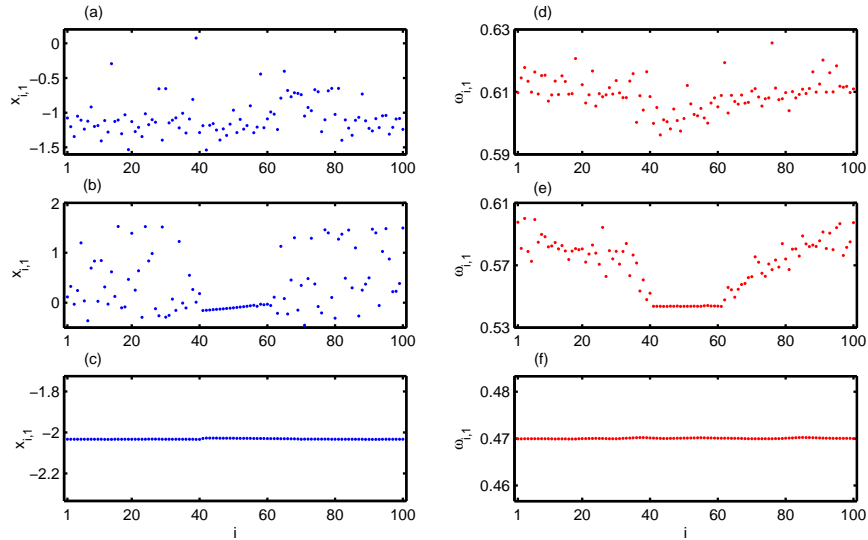,width=13.6cm}}
\caption{Left panels show snapshots of membrane potentials for (a) disordered state at $K_{ch} = 0.43$, (b) chimera state at $K_{ch} = 0.73$, and (c) coherent state at $K_{ch} = 1.10$. Right panels (d), (e), and (f) show the corresponding mean angular frequencies of disordered, chimera and coherent states. The homogeneous inter-later synaptic coupling delay-time $\tau=0.4$.}
\label{snap_homo_delay}
\end{figure*}

As mentioned earlier that the inter-layer information transmission may not be instantaneous, in general, and so the time-delay in this process should not be neglected.  Previous notable works include enhancement of synchrony in a network of Hindmarsh-Rose neuronal oscillators with time-delayed coupling \cite{delayprl}. Influence of time delay in the context of control of synchronization is studied in coupled excitable neurons in \cite{delayscholl}. Impact of information transmission delay on the synchronization transitions of modular networks in presence of both electrical and chemical synapses has also been studied in \cite{delayhybrid}. Here our aim is to analyze how the dynamical state in the upper layer varies in presence of time-delay in the inter-layer chemical synapses. At first we take the homogeneous inter-layer delay, i.e. the delay in the information transmission from upper to lower layer and from lower to upper layer are same i.e. $\tau_1=\tau_2=\tau$. Keeping $\tau$ fixed at a certain value and varying $K_{ch}$ we observe chimera patterns again as a link between incoherence and coherence as before. In Fig.~\ref{snap_homo_delay}, we fix the inter-layer chemical synaptic coupling delay as $\tau=0.4$ and vary the inter-layer chemical synaptic coupling strength $K_{ch}$. Here again for small values of $K_{ch}$, the upper layer remains disordered but as it increases, chimera pattern arises and sustains for a longer range of $K_{ch}$ compared to the previous case of instantaneous inter-layer coupling. Further increase in $K_{ch}$ gives rise to coherent dynamics in the layer. In the left panel of Fig. 5, the snapshot of amplitudes (membrane potential) depicting disordered, chimera and coherent states for $K_{ch}=0.43, 0.73$ and $1.10$ are shown in Figs.~\ref{snap_homo_delay}(a-c) respectively. Corresponding mean angular frequencies for disordered, chimera and coherent states are given in Figs.~\ref{snap_homo_delay}(d-f).

\begin{figure}
\centerline{\epsfig{file=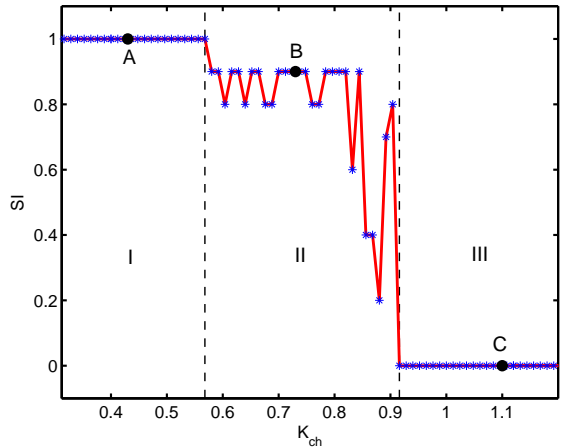,width=8.5cm}}
\caption{Strength of incoherence is plotted against inter-layer chemical synaptic coupling strength $K_{ch}$ for $\tau=0.4$. Regions I, II and III respectively stand for disordered, chimera and coherent states. The values pointed as A (at $K_{ch}=0.43$), B (at $K_{ch}=0.73$) and C (at $K_{ch}=1.10$) correspond to the exemplary snapshots given in Fig.~\ref{snap_homo_delay}(a), (b) and (c) respectively.}
\label{SI_homo_delay}
\end{figure}

As a characterization of the chimera states, we plotted the strength of incoherence SI, for different values of $K_{ch}$. As in Fig.~\ref{SI_homo_delay}, the upper layer remains disordered in the region I=$\{K_{ch}: 0.4\le K_{ch}<0.57\}$ with SI=1, chimera pattern emerges in the region II=$\{K_{ch}: 0.57\le K_{ch}\le 0.92\}$ where $0<$ SI $<1$  and the layer gets ordered if we further increase the value of $K_{ch}$ as the value of SI turns into zero.

\begin{figure}
\centerline{\epsfig{file=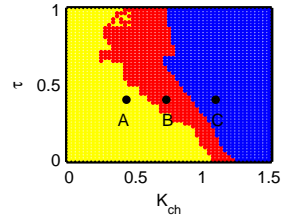,width=3.9cm}}
\caption{Two parameter phase diagram in the plane $K_{ch}-\tau$ of isolated neurons in the upper layer. Strength of incoherence is used as a measure for incoherence, coherence, and chimera states. Blue, red and yellow colors represent the region of coherent, chimera, and incoherent states respectively. Points A ($\tau=0.4$, $K_{ch}=0.43$), B ($\tau=0.4$, $K_{ch}=0.73$), C ($\tau=0.4$, $K_{ch}=1.10$) correspond to the values used in  Fig.~\ref{snap_homo_delay}(a), (b), (c) respectively.}
\label{twopara_homo}
\end{figure}

To get the complete understanding of the simultaneous effect of $K_{ch}$ and $\tau$, we rigorously plot the states of the upper layer network in the $K_{ch}-\tau$ parameter space for the range $K_{ch} \in [0,1.5]$ and $\tau \in [0,1.0]$ in Fig.~\ref{twopara_homo}. The strength of incoherence (SI) is used to distinguish between different dynamical states by changing  $K_{ch}$ and $\tau$ simultaneously. Blue, red and yellow colors stand for the region of coherent, chimera, and incoherent states respectively.  As $K_{ch}$ increases, the successive scenario of incoherent state, chimera state followed by coherent states remains unaltered for almost all the values of $\tau$ in the parameter space. The widening of the region reflecting chimera pattern due to the introduction of time delay $\tau$ is observed in the parameter space as well. These proves that the chimera patterns persist as a natural link between incoherence and coherence even in the presence of homogeneous delay in the inter-layer interaction, irrespective of the amount of time-delay. Also we must note that for the case of instantaneous inter-layer chemical synaptic coupling, Fig.~\ref{SIvary} shows chimera state starts occurring only when $K_{ch} \ge 1.075$. But the introduction of the information transmission delay $\tau$ in the inter-layer coupling brings about chimera pattern even when $K_{ch}$ is much smaller than $1.075$ which is clear from Fig.~\ref{twopara_homo}.

\subsection*{Heterogeneous inter-layer chemical synaptic coupling delays}

\begin{figure*}
\centerline{\epsfig{file=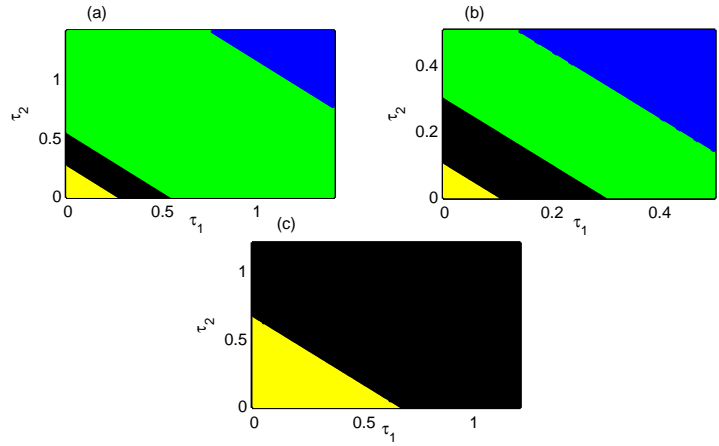,width=12cm}}
\caption{$(\tau_1,\tau_2)$ parameter space for N = 100 isolated identical Hindmarsh-Rose oscillators with (a) $K_{ch}=0.8$, (b) $K_{ch}=1.0$ and (c) $K_{ch}=0.35$. Strength of incoherence (SI) is used as a measure for incoherence, coherence, and chimera states. Blue region is for coherent, green is for coexistence of chimera and coherent, black is for coexisting chimera and incoherent, and yellow is for incoherent states.}
\label{twopara_hetero}
\end{figure*}

Next we consider the most general form of interaction between the two layers of oscillators by taking heterogeneous delays in the inter-layer synaptic coupling.  Formerly impact of heterogeneous time delay is investigated in the context of different types of cluster synchronization \cite{het1,het2}, synchrony \cite{het3} and amplitude death \cite{het4} of coupled neural networks. Information transmission delays from upper to lower layer i.e., $\tau_1$ and from lower to upper layer i.e., $\tau_2$ are different in this case. These heterogeneous time delays can also be reduced to homogeneous time delay $\tau=\frac{\tau_1+\tau_2}{2}$,  by a time shift transformation, recently proposed by Lucken et al. \cite{het5}. But our main concern will be with the emergence of chimera patterns due to the co-action of these different time-delays $\tau_1$ and $\tau_2$.

In order to do that, first we keep $K_{ch}$ fixed at $0.8$, where chimera was observed for a long range of $\tau$ in the homogeneous delay case (Fig.~\ref{twopara_homo}). We plot the $\tau_1-\tau_2$ parameter space for the range $\tau_1 \in [0,1.4]$ and $\tau_2 \in [0,1.4]$ in Fig.~\ref{twopara_hetero}(a). Initially, as $\tau_1$ and $\tau_2$ increase simultaneously, from the state of incoherence the upper layer network may achieve chimera state when $\tau_1$ and $\tau_2$ passes a certain value (satisfying $\tau_1+\tau_2 \gtrsim 0.26$). In fact, then we found a region (in black color) where incoherent and chimera pattern coexist in the parameter space. For further increment in $\tau_1$ and $\tau_2$, coexistence of chimera and coherence is observed (green region) and even higher values of $\tau_1$ and $\tau_2$ leads the upper layer to the state of coherence (blue region). The more the value of $\tau_1$, less the value of $\tau_2$ is needed (and vice versa ) for the network to attain coherent state as in Fig.~\ref{twopara_hetero}(a). This is how in this case, chimera may be found in a wide range of the $\tau_1-\tau_2$ parameter space.
	\par Next, taking $K_{ch}$ from the regime where synchrony is observed for a long range of $\tau$ (the homogeneous delay), we again discover a convincing enough chimera region in the $\tau_1-\tau_2$ parameter space followed by a synchronous region. For fixed $K_{ch}=1.0$, in the $\tau_1-\tau_2$ parameter space with $\tau_1 \in [0,0.5]$ and $\tau_2 \in [0,0.5]$, disordered state (yellow region) appears only for very small values of $\tau_1, \tau_2$ as shown in Fig.~\ref{twopara_hetero}(b). But as both the delays increase, the region of coexistence of incoherence and chimera (in black color) and a region of coexistence of coherence and chimera pattern (in green color) develops as in the previous case followed by the coherent state (blue region).

Finally, we fix $K_{ch}=0.35$ at such a value where no chimera was observed for almost any value of the homogeneous time-delay $\tau$ (Fig.~\ref{twopara_homo}). Taking $K_{ch}=0.35$, for $\tau_1$ and $\tau_2$ satisfying $\tau_1+\tau_2 \gtrsim 0.66$, the chimera state can emerge in the uncoupled neurons as in Fig.~\ref{twopara_hetero}(c). The region (in black color) of coexistence of disordered and chimera states is identified. But coherence has not been observed for almost any value of $\tau_1$ and $\tau_2$ in this case . In fact, even if we fix $K_{ch}$ at other smaller or larger values than the above values taken, we will have the similar type of $\tau_1-\tau_2$ space except only the range of $\tau_1$ and $\tau_2$ at which disordered, chimera and coherent states show up. This is how the information transmission delays can play a crucial role as far  as the formation of chimera pattern is concerned.  By introducing heterogeneous delay, we tried to make our model as general as possible, however, the effect of both $\tau_1$ and $\tau_2$ are similar as seen in Fig.~\ref{twopara_hetero}.

\section*{Discussion}
In summary, we have inspected how multi-layering can bring about chimera states in a network of uncoupled neurons where the multi-layering layer (lower layer)  of neurons are globally coupled. The neurons in the multi-layering layer has been assumed to be connected with electrical synapses whereas inter layer connection has been supposed to be of chemical synaptic type. The coaction of these two types of synapses leads the uncoupled layer (upper layer) to a chimera state. We discussed the existence of density dependent threshold for the emergence of chimera states in uncoupled neurons.  It is identified that delay in the inter layer coupling may enlarge the range of inter layer coupling strength for which the chimera pattern appears, compared to instantaneous inter layer coupling. We also obtained chimera states when we took other types of coupling in the common medium i.e. the multi-layering layer (see Supplementary Information). Our results therefore seem to be relevant for brain dynamics where coexistence of coherent and incoherent behaviors of the neurons appear.

\section*{Methods}
In the present work our object is to study the behavior of the uncoupled neurons which are not directly coupled rather they are communicated with each others via a common medium. The neuronal dynamics can be controlled by the coaction of two synapses, namely of electric and chemical synapses.

\subsection*{Hindmarsh-Rose neuronal model}
We consider each neuron in the multilayer network with Hindmarsh-Rose neuronal model dynamics. The Hindmarsh-Rose neuronal model is a popular for its chaotic bursting behavior and the original form is as follows:
\begin{equation}
\begin{array}{lcl}
\dot x=y+ax^2-x^3-z+I,\\
\dot y=1-dx^2-y,\\
\dot z=c(b(x-x_0)-z),
\end{array}
\end{equation}
where $x$-variable represents the membrane potential, $y$ and $z$ represent the transport of ions across the membrane through the fast and slow channels respectively. The variable $z$ corresponds the controls of speed of variation of the slow current and this speed is control by the small parameter $c$. Here the parameter $I$ denotes an external current that enters the neuron and $x_0$ controls delaying and advancing the activation of the slow current in the modeled neuron. For the sake of simplicity, after parameter redefinition or linear transformation \cite{transform} $x\rightarrow x, y \rightarrow 1-y, z \rightarrow  1+I+z, d\rightarrow a+\alpha, e\rightarrow -1-I-b x_0$, the above Eq. (1) can be rewritten as
\begin{equation}
\begin{array}{lcl}
\dot{x}=ax^2-x^3-y-z,\\
 \dot{y}=(a+\alpha)x^2-y,\\
\dot{z}=c(bx-z+e).
\end{array}
\end{equation}
The transformed model (2) is a phenomenological model that gives all the common dynamical features in a number of biophysical modeling studies of various bursting.

We consider Hindmarsh-Rose models as the nodes of the network (schematic diagram is shown in Fig.~\ref{schematic}) where both types of synapses (electrical and chemical) are present. The equations governing the dynamics in the upper and lower layer become
\begin{widetext}
\begin{equation}
\begin{array}{lcl}
\dot x_{i,1}=a{x_{i,1}}^2-{x_{i,1}}^3-y_{i,1}-z_{i,1}+K_{ch}(v_s-x_{i,1})\Gamma(x_{i,2}(t-\tau_2)),\\
\dot y_{i,1}=(a+\alpha){x_{i,1}}^2-y_{i,1},\\
\dot z_{i,1}=c(bx_{i,1}-z_{i,1}+e),
\end{array}
\end{equation}
\begin{equation}
\begin{array}{lcl}
\dot x_{i,2}=a{x_{i,2}}^2-{x_{i,2}}^3-y_{i,2}-z_{i,2}+K_{ch}(v_s-x_{i,2})\Gamma(x_{i,1}(t-\tau_1))+K_{el}\sum\limits_{j=1,j\neq{i}}^N E(x_{j,2}, x_{i,2}),\\
\dot y_{i,2}=(a+\alpha){x_{i,2}}^2-y_{i,2},\\
\dot z_{i,2}=c(bx_{i,2}-z_{i,2}+e),
\end{array}
\end{equation}
\end{widetext}
respectively, where $(x_{i,1},y_{i,1},z_{i,1})$ and $(x_{i,2},y_{i,2},z_{i,2})$ represent the state vectors for the neurons in the upper and lower layers respectively, $i=1, 2, \cdot \cdot \cdot ,N$; $N$ being the number of neurons in each of the layers of the network.  Here $\tau_1$ and $\tau_2$ are the time-delays required to propagate the information from upper to lower layer and lower to upper layer respectively.  The variables $x_{i,k}$ represent the membrane potentials, and
the variables $y_{i,k}$ and $z_{i,k}$ are the transport of ions across the membrane through the fast and slow channels, respectively for upper and lower layers for $k=1, 2$. We consider $c$ a small positive parameter so that $z_{i,k}$ varies much slower than $x_{i,k}$ and $y_{i,k}$ ($k=1,2$). The regular square-wave bursting is observed for the set of parameter values: $a = 2.8, \alpha = 1.6, c =0.001, b = 9$, and $e = 5$ (time series shown in Fig. S2(a) in the Supplementary Information). This system is monostable, that is, the coexistence of a stable equilibrium point and a limit
cycle has not been observed for this set of parameter values. The synapses are excitatory or inhibitory for the reversal potential $v_s$ greater or less than $x_{i,k}(t)$ for all $x_{i,k}(t)$ and all times $t$.
If $i$-th and $j$-th neurons are connected through electrical synapses then $E(x_{j,2}, x_{i,2})=x_{j,2}- x_{i,2}, \; i,j=1,2,...,N$. From physicist's perspective, at electrical synapse, gap junction between two neurons allows electron to move from one to another neuron via intercellular channels. This synapse is bidirectional and of a local character and occurring between those neurons which are spatially very close. By the mutual interaction through these synapses, neurons exhibit coherence or phase synchronization very easily and resulting into a group of synchronized neurons. The coupling strength associated with these synapses is $K_{el}$. Whereas, the chemical synaptic coupling function $\Gamma(x)$ is modeled by the sigmoidal nonlinear input-output function as
\begin{equation}
\begin{array}{lcl}
\Gamma(x)=\frac{1}{1+e^{-\lambda(x-\Theta_s)}},
\end{array}
\end{equation}
with $\lambda$ determining the slope of the function and $\Theta_s$ is the synaptic threshold. There is no such intercellular continuity at chemical synapses and no direct flow of electron from one neuron to another. The space gap between presynaptic and postsynaptic neurons is substantially greater at chemical synapses than electric synapses. Synaptic current flows from presynaptic neuron to postsynaptic neuron only in response to the secretion of neurotransmitters (e.g. acetylchosine, glutamate etc.). This synapse is either excitatory or inhibitory that depends on the neurotransmitters. We assume the chemical synapses are in excitatory for $v_s=2.0$ as it has a important function in information processing within the brain and throughout the peripheral nervous system.  We choose the threshold $\Theta_s=-0.25$ so as to make every spike in the isolated neuron burst to reach the threshold and we fix the value $\lambda=10$. Here $K_{ch}$ is the coupling strength associated with the chemical synapses. We assume, for simplicity, both the synapses, namely electrical and chemical synapses transmit the electron bidirectionally from one to another neuron. The systems (3) and (4) are integrated using fifth-order Runge-Kutta-Fehlberg integration scheme with integration time step $0.01$ for non-delayed cases i.e. $\tau_1=\tau_2=0.0.$ In presence of synaptic coupling delay, we integrate systems (3) and (4) using modified Heun method with integration time step $0.01.$

\subsection*{Characterization of chimera state: strength of incoherence}
To characterize the disordered, chimera and coherent states, we use the statistical measures using the time series of the network \cite{SImeasure}. In order to do that we measure the strength of incoherence (SI) using a local standard deviation analysis. To calculate SI, we firstly define the transformations $w_{i,1}=x_{i,1}-x_{i+1,1}$, $i=1, 2, \cdot \cdot \cdot ,N$. We divided the total number of neurons in upper layer into $M$ (even) bins of equal length $n=N/M$ and $\sigma_1(m)$ is the local standard deviation in each of these bins as follows
\begin{equation}
\begin{array}{lcl}
\sigma_1(m)=\left \langle \sqrt{\frac{1}{n} \sum\limits_{j=n(m-1)+1}^{mn} [w_{j,1}-\left \langle w_1\right\rangle]^2}~~\right\rangle_t,
\end{array}
\end{equation}
with $\left \langle w_1 \right\rangle=\frac{1}{N} \sum\limits_{i=1}^N w_{i,1}(t)$;  $m=1, 2, \cdot \cdot \cdot ,M$. Then SI is defined as
\begin{equation}
\begin{array}{lcl}
\mbox{SI} =1-\frac{\sum\limits_{m=1}^M s_m}{M}, ~~~~~~~~~~~~ s_m=\Theta(\delta-\sigma_1(m)),
\end{array}
\end{equation}
where $\Theta(\cdot)$ is the Heaviside step function and $\delta$ is a predefined threshold.  Consequently, the values SI$=1$, SI$=0$ and $0<$ SI $<1$ correspond to the incoherent, coherent and chimera states respectively.

\begin{acknowledgments}
This research was supported by the Slovenian Research Agency (Grants P5-0027 and J1-7009).
\end{acknowledgments}

\clearpage

\setcounter{page}{1}
\renewcommand\thepage{\roman{page}}

\setcounter{equation}{0}
\setcounter{figure}{0}
\setcounter{table}{0}
\makeatletter
\renewcommand{\theequation}{S\arabic{equation}}
\renewcommand{\thefigure}{S\arabic{figure}}
\renewcommand{\bibnumfmt}[1]{[S#1]}
\renewcommand{\citenumfont}[1]{S#1}

\onecolumngrid
\begin{center}
\begin{large}
\textbf{Chimera states in uncoupled neurons induced by a multilayer structure\\Supplementary information}
\end{large}
\end{center}

\twocolumngrid

We first briefly discuss the formation of chimera state in the upper layer of uncoupled neurons in the presence of non-local interactions among the neurons of the lower layer (the multi-layering layer). For this purpose, we consider non-local interaction in the lower layer where both types of synapses (electrical and chemical) are present as proposed in the article. With instantaneous inter layer chemical synaptic coupling, the equations governing the dynamics in the upper and lower layer are
\begin{widetext}
\begin{equation}
\begin{array}{lcl}
\dot x_{i,1}=a{x_{i,1}}^2-{x_{i,1}}^3-y_{i,1}-z_{i,1}+K_{ch}(v_s-x_{i,1})\Gamma(x_{i,2}(t)),\\
\dot y_{i,1}=(a+\alpha){x_{i,1}}^2-y_{i,1},\\
\dot z_{i,1}=c(bx_{i,1}-z_{i,1}+e),
\end{array}
\end{equation}
\begin{equation}
\begin{array}{lcl}
\dot x_{i,2}=a{x_{i,2}}^2-{x_{i,2}}^3-y_{i,2}-z_{i,2}+K_{ch}(v_s-x_{i,2})\Gamma(x_{i,1}(t))+K_{el}\sum\limits_{j=i-P,j\neq{i}}^{i+P} E(x_{j,2}, x_{i,2}),\\
\dot y_{i,2}=(a+\alpha){x_{i,2}}^2-y_{i,2},\\
\dot z_{i,2}=c(bx_{i,2}-z_{i,2}+e),
\end{array}
\end{equation}
\end{widetext}
respectively. Square-wave bursting dynamics is assumed for all the nodes of the two layers with the set of parameter values: $a = 2.8, \alpha = 1.6, c =0.001, b = 9$, and $e = 5$. The chemical synaptic coupling function  $\Gamma(x)$ is the following
\begin{equation}
\begin{array}{lcl}
\Gamma(x)=\frac{1}{1+e^{-\lambda(x-\Theta_s)}},
\end{array}
\end{equation}
whereas $E(x_{j,2}, x_{i,2})=x_{j,2}- x_{i,2}, \; i,j=1,2,...,N$ with $v_s=2.0$, $\Theta_s=-0.25$ and $\lambda=10$, as before. Here $P$ is the number of coupled nearest neighbor neurons on both sides on a ring.

\begin{figure*}
\centerline{\epsfig{file=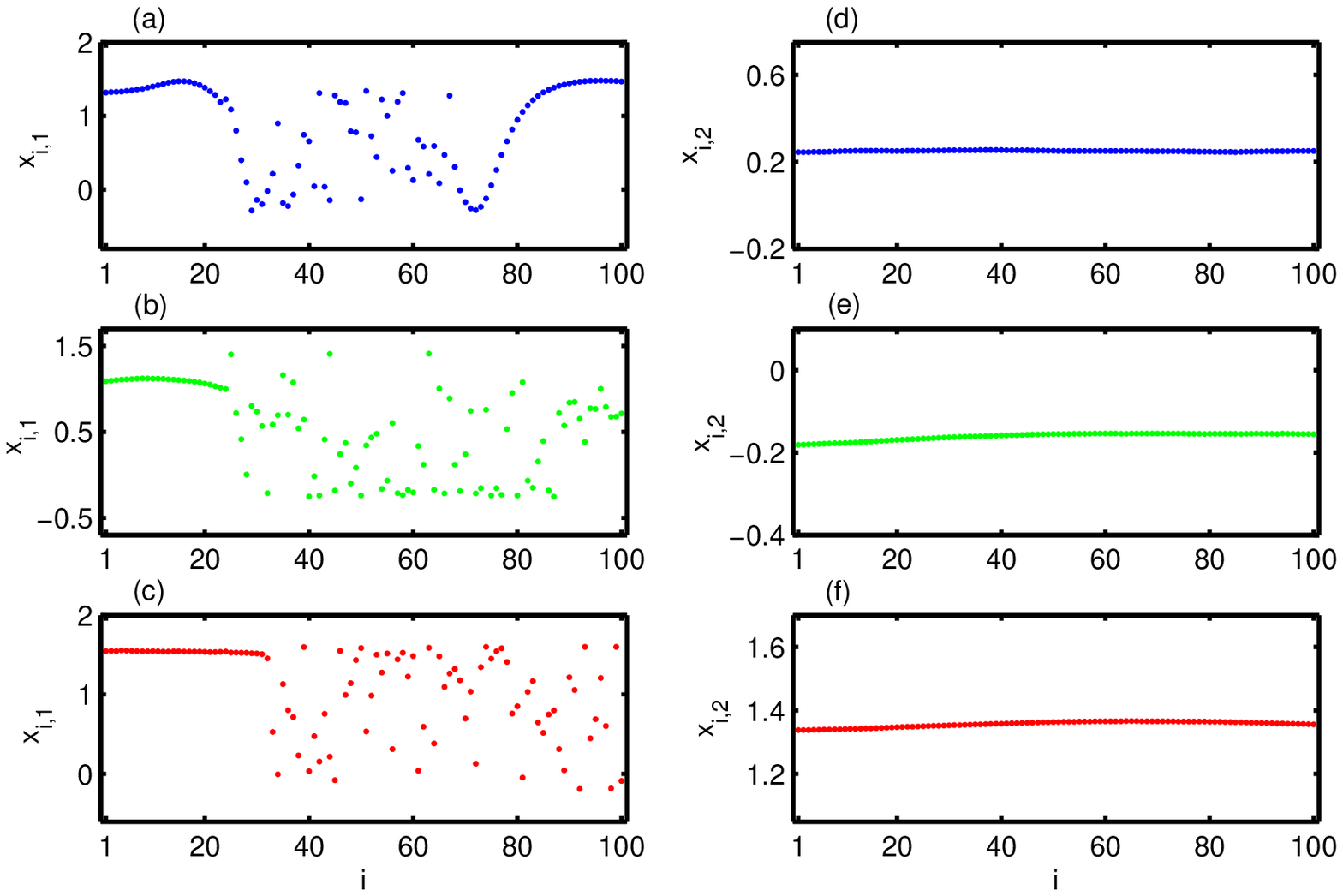,width=13cm}}
\caption{Left panels show the snapshots of membrane potentials $x_{i, 1}$ of the upper layer neurons $(i=1,2,...,100)$ depicting chimera state for (a) $P=10$ and $K_{ch}=0.9$, (b) $P=20$ and $K_{ch}=1.1$, and (c) $P=30$ and $K_{ch}=1.1$. Right panels show the snapshots for the corresponding membrane potentials $x_{i, 2}$ of the lower layer neurons in (d), (e) and (f) respectively. Here $K_{el}=2.0.$.}
\label{snapshot_nolocal}
\end{figure*}

At first, we fix the electrical synaptic coupling strength $K_{el}=2.0$ and vary the number of nearest neighbors $P$ and chemical synaptic coupling strength $K_{ch}$ to observe the behavior of uncoupled neurons. Taking $P=10$ for which with $K_{ch}=0.0$, the lower layer of neurons interacting through electrical coupling gets synchronized. Then switching $K_{ch}$ on, we observe chimera state in the upper layer of uncoupled neurons. Typical snapshots for both the upper and lower layer neuron's membrane potentials with $K_{ch}=0.9$ are shown in Figs.~\ref{snapshot_nolocal}(a) and (d) respectively. Then we make $P=20$ with the lower layer neurons in coherent state. For $K_{ch}=1.1$, chimera pattern is seen in the upper layer of neurons. Snapshots of membrane potentials of both the layers are given in Figs.~\ref{snapshot_nolocal}(b) and (e) respectively. Finally, taking $P=30$, we detect chimera state in the upper layer while the lower layer is in coherent state. Snapshots for $K_{ch}=1.1$ are shown respectively in Figs.~\ref{snapshot_nolocal}(c) and (f).

\begin{figure*}
\centerline{\epsfig{file=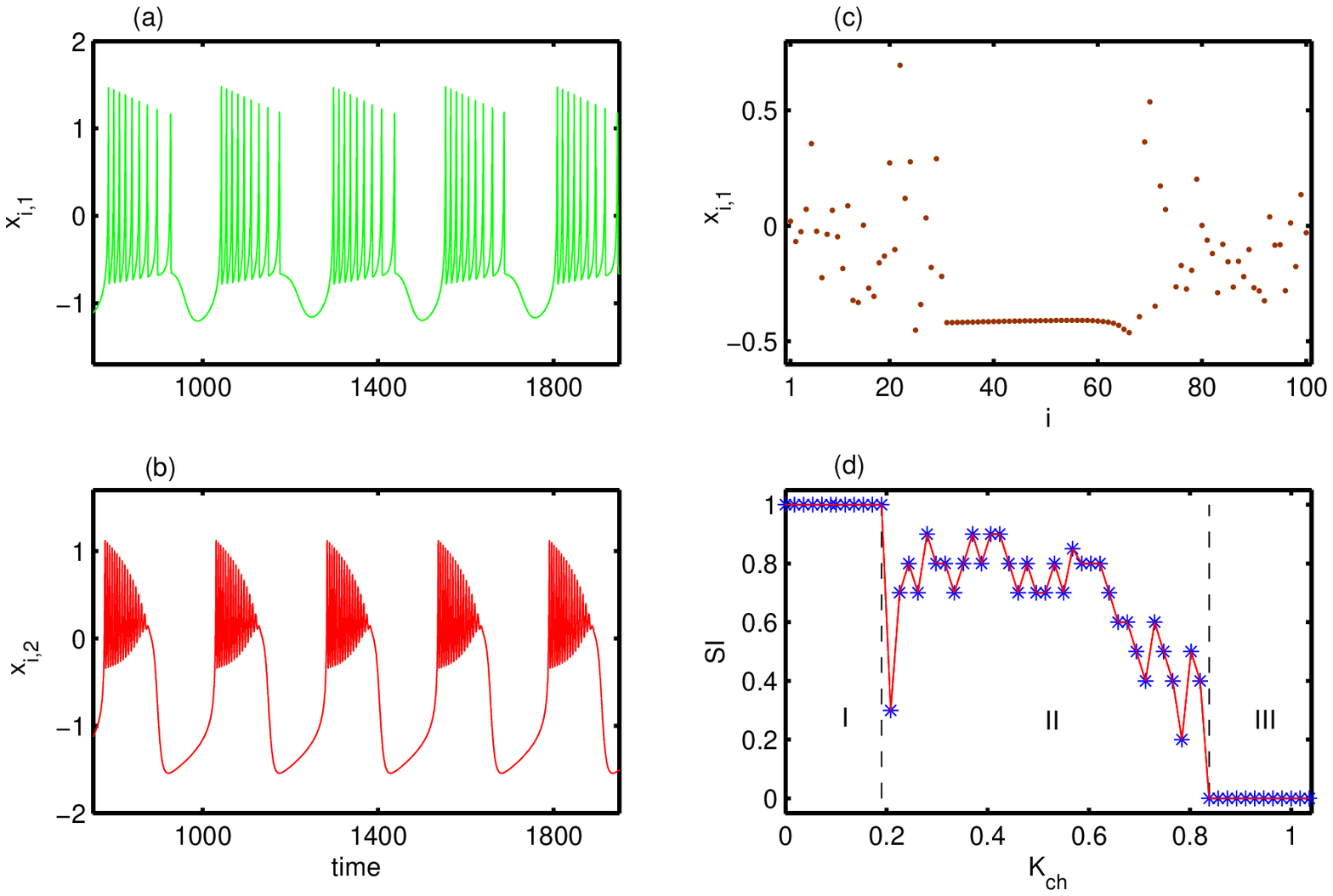,width=13cm}}
\caption{Left panels show the time series of membrane potentials (a) $x_{i, 1}$ exhibiting square wave bursting (for $a_1=2.8$) and (b) $x_{i, 2}$ exhibiting plateau bursting (for $a_2=2.2$), $i=1$.  (c) Snapshot characterizing chimera state with $K_{el}=1.5$ and $K_{ch}=0.53$, (d) Variation of strength of incoherence (SI) by changing the inter-layer chemical synaptic coupling strength $K_{ch}$ with $K_{el}=1.5$.}
\label{paramis}
\end{figure*}

Secondly, we provide further details with regards to the chimera pattern in the proposed multi-layer structure in  presence of inter layer heterogeneity. In particular, we are concerned with two layers having different dynamical natures of the neurons by considering parameter mismatches. For this purpose, without loss of generality, we only change the parameter $a$ that appears in the first two equations of both Eqns. $(S1)$ and $(S2)$. Then the equations governing the dynamics in the upper and lower layer become
\begin{widetext}
\begin{equation}
\begin{array}{lcl}
\dot x_{i,1}=a_1{x_{i,1}}^2-{x_{i,1}}^3-y_{i,1}-z_{i,1}+K_{ch}(v_s-x_{i,1})\Gamma(x_{i,2}(t)),\\
\dot y_{i,1}=(a_1+\alpha){x_{i,1}}^2-y_{i,1},\\
\dot z_{i,1}=c(bx_{i,1}-z_{i,1}+e),
\end{array}
\end{equation}
\begin{equation}
\begin{array}{lcl}
\dot x_{i,2}=a_2{x_{i,2}}^2-{x_{i,2}}^3-y_{i,2}-z_{i,2}+K_{ch}(v_s-x_{i,2})\Gamma(x_{i,1}(t))+K_{el}\sum\limits_{j=1,j\neq{i}}^N E(x_{j,2}, x_{i,2}),\\
\dot y_{i,2}=(a_2+\alpha){x_{i,2}}^2-y_{i,2},\\
\dot z_{i,2}=c(bx_{i,2}-z_{i,2}+e)
\end{array}
\end{equation}
\end{widetext}
respectively. All the parameters except $a_1$ and $a_2$ are taken same as above. As discussed in the article, for $a_1=2.8$, the individual neurons of the upper layer show square wave bursting (shown in Fig.~\ref{paramis}(a)) and for $a_2=2.2$,  the lower layer neurons exhibit plateau bursting (Fig.~\ref{paramis}(b)), if no coupling is present.

\begin{figure*}
\centerline{\epsfig{file=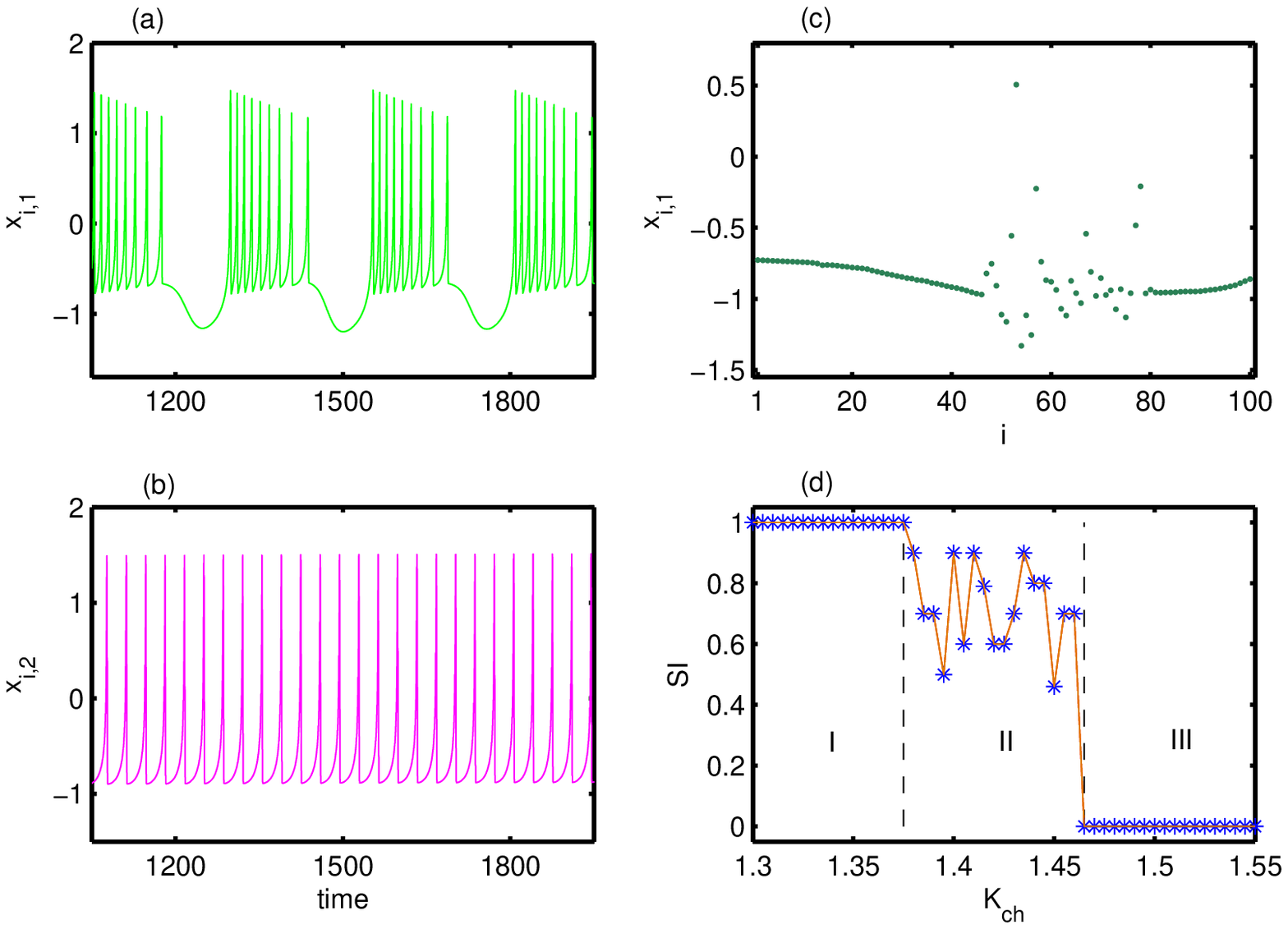,width=12.5cm}}
\caption{Left panels show the time series of membrane potentials (a) $x_{i, 1}$ exhibiting square wave bursting for $a_1=2.8$ and (b) $x_{i, 2}$ exhibiting spiking for $a_2=3.0$ (here $i=1$).  (c) Snapshot characterizing chimera state with $K_{el}=1.2$ and $K_{ch}=1.43$, (d) strength of incoherence (SI) depending on $K_{ch}$. The regions I, II and III stand for incoherent, chimera and coherent states respectively.}
\label{paramis2}
\end{figure*}

Keeping $K_{el}=1.5$ fixed and switching $K_{ch}$ on, we observe that after being disordered, the upper layer experiences chimera pattern followed by synchronized state for increasing values of $K_{ch}$. Snapshot characterizing chimera pattern for $K_{el}=1.5$ and $K_{ch}=0.53$ can be seen in Fig.~\ref{paramis}(c). Fig.~\ref{paramis}(d) depicts the variation of strength of incoherence (SI) depending on $K_{ch}$. The regions I, II and III represent incoherent, chimera and coherent states respectively.

Next, we consider $a_1=2.8$ as before and $a_2=3.0$. For this choice, in absence of any interaction between the neurons, the neurons of the upper layer show square wave bursting (shown in Fig.~\ref{paramis2}(a)) and the lower layer neurons exhibit spiking, as shown in Fig.~\ref{paramis2}(b).

For fixed $K_{el}=1.2$ and increasing $K_{ch}$, we observe that the upper layer experiences chimera pattern followed by synchronized state after being disordered initially. Snapshot depicting chimera state for $K_{el}=1.2$ and $K_{ch}=1.43$ is given in Fig.~\ref{paramis2}(c). Fig.~\ref{paramis2}(d) shows the variation of strength of incoherence (SI) depending on $K_{ch}$.

\end{document}